\documentclass[sigconf, screen]{acmart}
\renewcommand\footnotetextcopyrightpermission[1]{} 
\AtBeginDocument{%
  }

\setcopyright{acmlicensed}
\copyrightyear{2018}
\acmYear{2018}
\acmDOI{XXXXXXX.XXXXXXX}
\acmConference[Conference acronym 'XX]{Make sure to enter the correct
  conference title from your rights confirmation email}{June 03--05,
  2018}{Woodstock, NY}
\acmISBN{978-1-4503-XXXX-X/2018/06}

\acmSubmissionID{7662}

\usepackage[most]{tcolorbox}
\usepackage{enumitem}

\newtcolorbox{promptbox}[1]{
  enhanced jigsaw,
  breakable,
  colback=white,
  colframe=black,
  colbacktitle=black,
  coltitle=white,
  title={#1},
  title after break={#1\ (cont.)},
  boxrule=0.7pt,
  arc=2mm,
  left=1mm,
  right=1mm,
  top=0.8mm,
  bottom=0.8mm,
  fonttitle=\bfseries,
  fontupper=\footnotesize,
  before skip=6pt,
  after skip=6pt,
  toprule at break=0.7pt,
  bottomrule at break=0.7pt,
  pad at break=1mm
}
\usepackage{multirow}
\begin{document}

\title{Administrative Decentralization in Edge-Cloud Multi-Agent for Mobile Automation}

\author{Senyao Li}
\authornote{Both authors contributed equally to this research.} 
\email{D202581873@hust.edu.cn}
\affiliation{%
  \institution{School of Computer Science and Technology, Huazhong University of Science and Technology}
  \country{China} 
}

\author{Zhigang Zuo}
\authornotemark[1] 
\affiliation{%
  \institution{School of Computer Science and Technology, Huazhong University of Science and Technology}
  \country{China}
}

\author{Haozhao Wang}
\authornote{Corresponding author.} 
\email{haozhaowang@hust.edu.cn}
\affiliation{%
  \institution{School of Computer Science and Technology, Huazhong University of Science and Technology}
  \country{China}
}

\author{Junyu Chen}
\affiliation{%
  \institution{School of Computer Science and Technology, Huazhong University of Science and Technology}
  \country{China}
}

\author{Zhanbo Jin}
\affiliation{%
 \institution{International School Beijing University of Posts and Telccommunication}
   \country{China}
 }

\author{Ruixuan Li}
\affiliation{%
  \institution{School of Computer Science and Technology, Huazhong University of Science and Technology}
    \country{China}
}

\renewcommand{\shortauthors}{Trovato et al.}

\begin{abstract}
  Collaborative edge-cloud frameworks have emerged as the mainstream paradigm for mobile automation, mitigating the latency and privacy risks inherent to monolithic cloud agents.
However, existing approaches centralize administration in the cloud while relegating the device to passive execution, inducing a cognitive lag regarding real-time UI dynamics.
To tackle this, we introduce \textbf{AdecPilot} by applying the principle of administrative decentralization to the edge-cloud multi-agent framework, which redefines edge agency by decoupling high-level strategic designing from tactical grounding. AdecPilot integrates a UI-agnostic cloud designer generating abstract milestones with a bimodal edge team capable of autonomous tactical planning and self-correction without cloud intervention.
Furthermore, AdecPilot employs a Hierarchical Implicit Termination protocol to enforce deterministic stops and prevent post-completion hallucinations.
Extensive experiments demonstrate proposed approach improves task success rate by 21.7\% while reducing cloud token consumption by 37.5\% against EcoAgent and decreasing end to end latency by 88.9\% against CORE. The source code is available at \url{https://anonymous.4open.science/r/Anonymous_code-B8AB}.
\end{abstract}

\maketitle

\section{Introduction}

Mobile platforms have evolved into the primary interface for digital life, catalyzing the rise of LLM-driven autonomous agents for mobile automation~\cite{chen2025multi, DBLP:conf/icml/ZhangY0LHZL0W0W25,fan2025core,aaai/WeiD0L025}. However, current architectures~\cite{liu2025promptr1collaborativeautomaticprompting} face a fundamental dilemma: monolithic cloud controllers~\cite{chen2025multi} incur prohibitive latency and privacy risks, while strict on-device~\cite{DBLP:conf/kdd/00010Y00025, DBLP:conf/icml/ZhangNF00025} constraints hinder the scalability of domain-specific models~\cite{ICLR2025_01a83bc2,han2025swarm}. To address this dilemma, the edge-cloud collaborative paradigm~\cite{DBLP:conf/www/ShaoHLX25}, which deploys a capable agent in the cloud and a lightweight agent at the edge to collaboratively implement the mobile automation, is emerging as the mainstream approach~\cite{aaai/HuTXDCFK25,li2025collaborative}.

\begin{figure}[t]
    \centering
    \includegraphics[width=\linewidth]{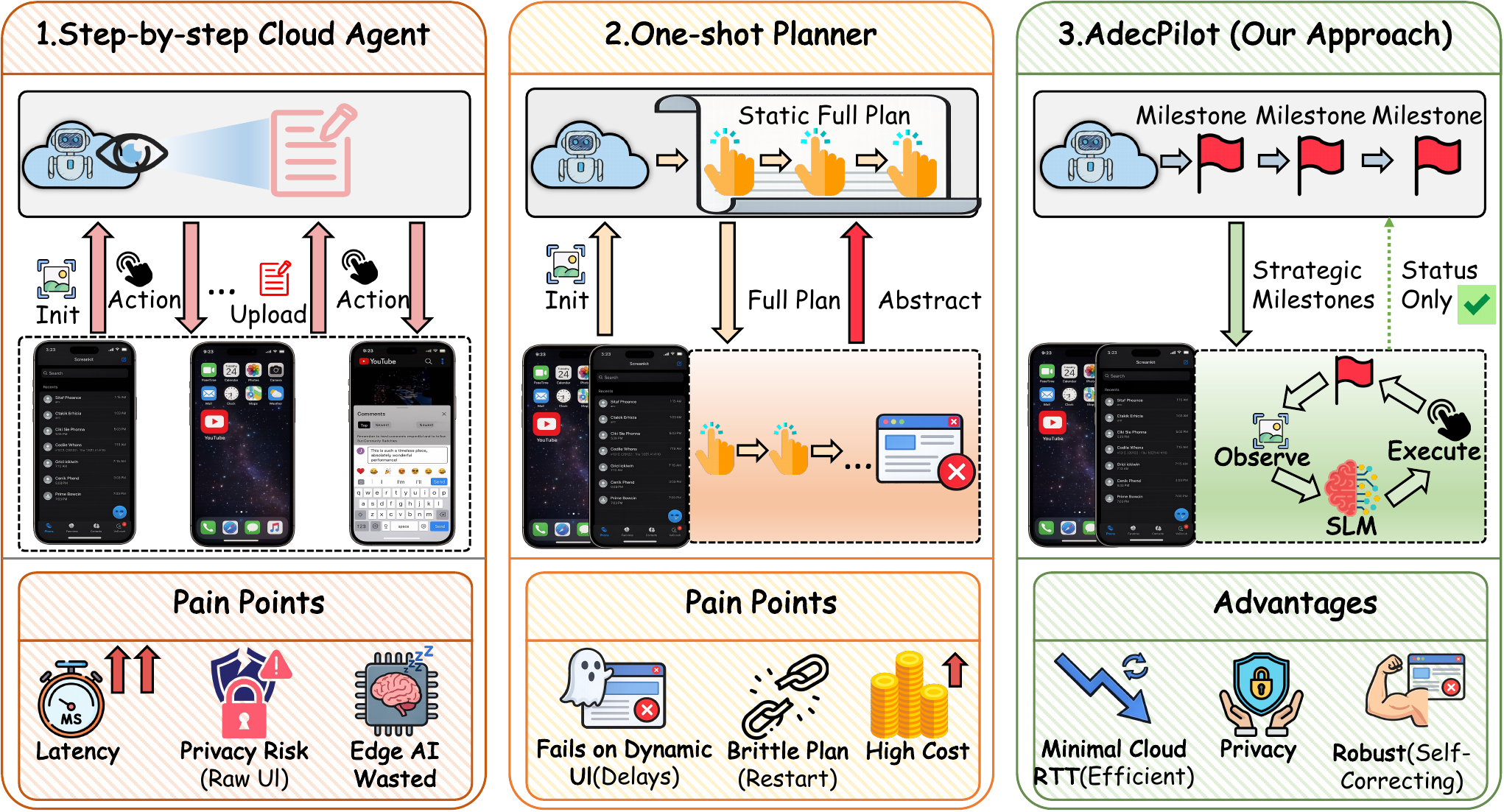}
    \caption{\textbf{Evolution of Mobile Agent Paradigms.} 
    \textit{Left \& Middle:} Conventional methods struggle with high latency (1) or brittle static plans failing under perturbations (2). 
    \textit{Right:} \textsc{AdecPilot} (3) decouples Strategic Milestones from Tactical Grounding. Autonomous edge \textbf{Planning} ensures robust local resolution, minimal privacy exposure, and minimal cloud consumption.}
    \label{fig:motivation}
\end{figure}

However, edge-cloud collaboration suffers from a substantial challenge, i.e.,  reconciling the tension between two fundamental principles:\\
\noindent$\bullet$ The \textit{scaling law of intelligence} posits that reasoning capabilities scale with model parameter counts, favoring large cloud models for complex logic.\\
\noindent$\bullet$ The \textit{law of observability} states that planning efficacy correlates directly with the fidelity and immediacy of real-time data, favoring edge access for precise execution. \\
To balance these two principles, existing works propose leveraging the cloud agent to perform strategic oversight while assigning the edge agent to handle execution~\cite{yi2025ecoagent}, which falls into two main categories. The first category~\cite{zhang2025appagent,long2025seeing,fan2025core} involves the cloud agent generating a short-term plan and iteratively refining it into a complete plan throughout the process based on feedback from the edge agent’s execution. The second category~\cite{yi2025ecoagent} entails the cloud agent producing a full plan upfront and then progressively correcting it over time in response to execution feedback from the edge agent. 
Although these methods have achieved considerable success, both rely entirely on the cloud side to handle all planning-related tasks, relegating the edge agent to a purely mechanical executor, thus resulting in \textit{an inability to plan against the visible real-time UI}.
Specifically, minor deviations such as unexpected icon relocations remain invisible to the remote planner, often causing the entire execution chain to fail before any error is even detected~\cite{DBLP:conf/icml/ChawlaVS023,zhan2025prism}.
Moreover, to avoid exposing sensitive user data, these approaches transmit only compressed summaries or cropped image patches, forcing the cloud-based planner to operate in a severely degraded visual environment~\cite{DBLP:conf/www/ShaoHLX25,huang2025cued}.
In summary, existing methods suffer from \textit{Remote Commander Paradox}: the entity endowed with the highest intelligence has the poorest perception of the current interface, while the edge observer who possesses real-time visual access remains incapable of performing planning.

To address the above challenges, we advocate for \textit{Administrative Decentralization} within the computational system architecture. We reimagine the cloud as a strategic leader responsible solely for sparse top-level design while delegating concrete planning and execution to the edge.
We propose \textsc{AdecPilot}.
This framework redefines the collaborative boundary by separating strategic intent from tactical implementation.
Adhering to the Scaling Law of Intelligence and Law of Observability, we preserve coarse-grained cloud supervision while empowering the edge with real-time UI observation and self-correction.
To address the modality mismatch between high-latency visual diagnosis and high-speed text execution, our edge visual agent performs local planning and observation while utilizing text agent for execution and correction.
The system transmits only specific actions rather than indiscriminate screen summaries strictly when the correction step limit is exceeded.

To operationalize this decentralization, \textsc{AdecPilot} integrates a UI-agnostic Cloud Strategic Director for high-level decomposition with a Tactical Edge Team comprising a Vision Orchestrator and Textual Executor.
Specifically, the cloud defines abstract milestones to guide the global trajectory while the edge autonomously resolves dynamic UI variances via local planning loops.
This architecture confines heavy visual processing and atomic decision-making to the device, effectively reducing cloud token consumption and ensuring privacy exposure minimization.
To safeguard execution, we further design the Hierarchical Implicit Termination protocol.
By restricting validation to the final milestone, this mechanism enforces a deterministic stop upon logic exhaustion and prevents post-completion hallucinations common in lightweight models.
Empirical results confirm that this architectural decoupling renders the system immune to network volatility and maintains baseline responsiveness even under severe bandwidth constraints where monolithic models fail. Compared to visual baselines like M3A~\cite{long2025seeing}, \textsc{AdecPilot} achieves 388.7$\times$ uplink data reduction and 43.8$\times$ computational efficiency gain via trajectory distillation.

Our primary contributions are as follows:

\begin{itemize}
    \item \textbf{Redefining Edge Agency:} We propose \textsc{AdecPilot} to decouple strategic cloud milestones from autonomous edge planning. This hierarchical separation ensures robustness against dynamic UI perturbations while minimizing cloud dependency.
    
    \item \textbf{Bimodal Autonomy \& HIT Protocol:} We integrate a VLM-based Orchestrator and Text-based Executor for local planning without cloud intervention. Additionally, our Hierarchical Implicit Termination (HIT) protocol enforces deterministic zero latency exits, effectively preventing post-completion hallucinations.
    
    \item \textbf{SOTA Efficiency:} Extensive evaluations demonstrate that \textsc{AdecPilot} significantly outperforms state-of-the-art baselines, validating its superiority in reducing cloud token consumption and overcoming transmission latency bottlenecks.
\end{itemize}

\section{Related Work}
\label{sec:related_work}

\subsection{Cloud Agents for UI Automation.}
Advent of Multimodal Large Language Models (MLLMs) transitioned UI automation from heuristic scripts to vision-driven agents~\cite{DBLP:conf/euromlsys/AslanidisKLC25,DBLP:conf/nips/BaiZPCSLK24}. Pioneering frameworks, including AppAgent~\cite{zhang2025appagent} and T3A~\cite{xu2025androidlab}, adopt stepwise execution paradigms. They utilize monolithic cloud MLLMs to process full-resolution screenshots for atomic action generation~\cite{xu2025androidlab}. Recent efforts like PRISM incorporate video history to capture temporal execution context~\cite{zhan2025prism}, attempting to resolve short-term memory deficits inherent to static screenshot analysis. While demonstrating competitive success rates on constrained benchmarks like AndroidWorld~\cite{ICLR2025_01a83bc2}, these cloud-centric paradigms suffer from fundamental architectural flaws. Primarily, continuous transmission of raw pixels induces a prohibitive trilemma: excessive token consumption, unacceptable network latency, and severe visual privacy leakage~\cite{aaai/WeiD0L025}. Furthermore, cloud planners operate without real-time state perception, leading to inevitable semantic mismatch when confronting dynamic UI mutations.

\subsection{Collaborative Edge-Cloud Multi-Agent}
Mitigating resource constraints, recent research pivots toward hybrid collaborative architectures~\cite{aaai/WeiD0L025,icml/Guo0CLHL025,rao2025multi}. General-purpose frameworks including AdaSwitch~\cite{DBLP:conf/emnlp/0015WCWFWWZY24} and Division-of-Thoughts~\cite{DBLP:conf/www/ShaoHLX25} explore adaptive mechanisms dynamically distributing inference loads across heterogeneous models based on sample difficulty~\cite{icml/ClintonCZK25}. Within mobile domains, EcoAgent~\cite{yi2025ecoagent} minimizes cloud interaction via one-shot planning strategies, generating comprehensive action sequences upfront~\cite{DBLP:conf/www/Cheng0W0QL0RM23}. Conversely, CORE~\cite{fan2025core} addresses privacy concerns by restricting transmission to text-based UI representations. Despite reducing cloud dependency, these approaches exhibit fundamental architectural flaws in dynamic GUI environments. Such frameworks treat edge modules as passive actuators lacking autonomous tactical resolution. Furthermore, text-only transmissions in CORE~\cite{fan2025core} discard crucial spatial constraints, resulting in structural ambiguity during execution. Ultimately, existing collaborative paradigms fail to achieve robust Intent Grounding. They physically distribute computation but fail to implement administrative decentralization.

\begin{figure}[t]
    \centering
    \includegraphics[width=\linewidth]{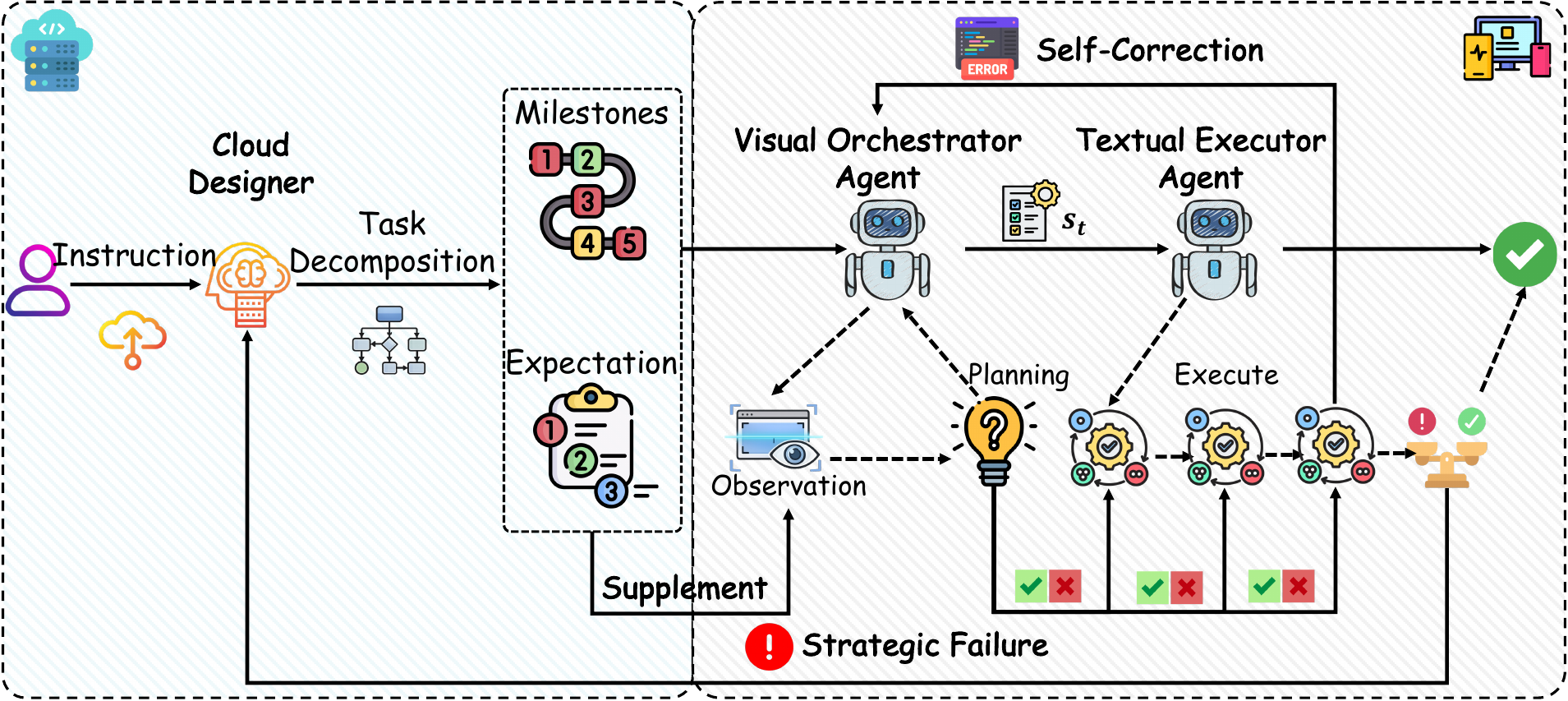}
    \vspace*{-10pt}
    \caption{\textbf{Overview of \textsc{AdecPilot}.} The UI-Agnostic Cloud Designer generates abstract milestones, while the Bimodal Edge Team autonomously executes them via local planning. This hierarchical loop enables real-time self-correction, ensuring robust and privacy-preserving automation.}
    \vspace*{-10pt}
\label{fig:framework_overview}
\end{figure}

\section{Problem Formulation}

We formalize multimodal mobile agent workflow as hierarchical decision process parameterized by action space $\mathcal{A}$, observation space $\mathcal{O}$, and synchronization cost function $\mathcal{C}_{sync}$. At step $t$, system operates based on defined inputs:

\noindent \textbf{Task Instruction} $L_{cmd} \in \mathcal{L}$: Natural language goal provided by user. \noindent \textbf{Application Metadata} $C_m$: Static invariant functional schema resolving app specific structural priors. \noindent \textbf{Strategic Decomposition}: Unlike monolithic oracles $f: \mathcal{L} \times \mathcal{O} \to \mathcal{A}$ depending upon continuous visual feedback, cloud designer $\Psi_c$ functions as open loop strategic designer. Initiating task, module maps user instruction $L_{cmd}$, initial empty history $\emptyset$, and abstract app metadata $C_m$ directly to sequence of UI agnostic milestones $G$, strictly bypassing real time observation $o_t$:
\begin{equation}
\label{eq:cloud_planning}
G = \{ (g_1, E_1), \dots, (g_K, E_K) \} \sim P_{\Psi_c}(\cdot \mid L_{cmd}, \emptyset, C_m)
\end{equation}
Tuple $(g_k, E_k)$ encapsulates strategic subgoal $g_k$ and expected visual invariant $E_k$, explicitly excluding low level UI directives. Formulation prioritizes stable business logic over transient rendering details. Bridging inference gap between abstract goals and concrete actions, bimodal edge pipeline assumes tactical autonomy. Process utilizes complementary sequential models: vision centric orchestrator $\Phi_m$ aligns visual observation $V_t$ with expected state $E_{k(t)}$ synthesizing meta instruction $s_t$, while text centric executor $\Lambda_e$ grounds $s_t$ against textual hierarchy $U_t$ yielding atomic action $a_t$. Execution pipeline is formally factorized exposing strict sequential dependency:
\begin{equation}
\label{eq:edge_execution}
s_t = \Phi_m(V_t,g_{k(t)}, E_{k(t)}), \quad a_t = \Lambda_e(s_t, U_t)
\end{equation}
where $g_{k(t)} \in G$ denotes currently active milestone at step $t$.

System descriptive formulation quantifies total cloud communication cost structure over task lifespan. Formulation establishes rigorous information bottleneck, enforcing administrative decentralization principle. Unlike stepwise generation where synchronization cost scales linearly with trajectory length and observation size, decomposition approach limits cloud interaction to initial planning phase and sparse replanning moments triggered by failure. Total synchronization cost $\mathcal{C}_{total}$ is formulated descriptively as sum of payload token volumes:
\begin{equation}
\label{eq:optimization}
\mathcal{C}_{total} = |L_{cmd}| + \sum_{k=1}^{K} \mathbb{I}(f_k) \cdot |H_{fail}^{(k)}|
\end{equation}
Here, operator $|\cdot|$ computes discrete token volume quantifying transmission bandwidth. Initial uplink payload comprises purely task instruction $L_{cmd}$. Variable $H_{fail}^{(k)} \in \mathcal{H}$ defines textual diagnostic payload encapsulating failed milestone $g_k$, expected invariant $E_k$, alongside error execution trajectory, transmitted specifically upon failure $f_k$. Transmitting solely textual payload $H_{fail}^{(k)}$ provides cloud designer sufficient replanning context while mitigating visual data leakage and minimizing synchronization overhead. Term $\mathbb{I}(f_k)$ denotes indicator function for failure event $f_k$, formally defined as:
\begin{equation}
\mathbb{I}(f_k) = \begin{cases}
1, & \text{if } \Lambda_e \text{ fails to reach expected state } E_k \text{ of } g_k \\
0, & \text{otherwise}
\end{cases}
\end{equation}
Formulation mathematically decouples strategic design from tactical execution, confining cumulative errors to local limits.

\begin{figure*}[t]
    \centering
    \includegraphics[width=\linewidth]{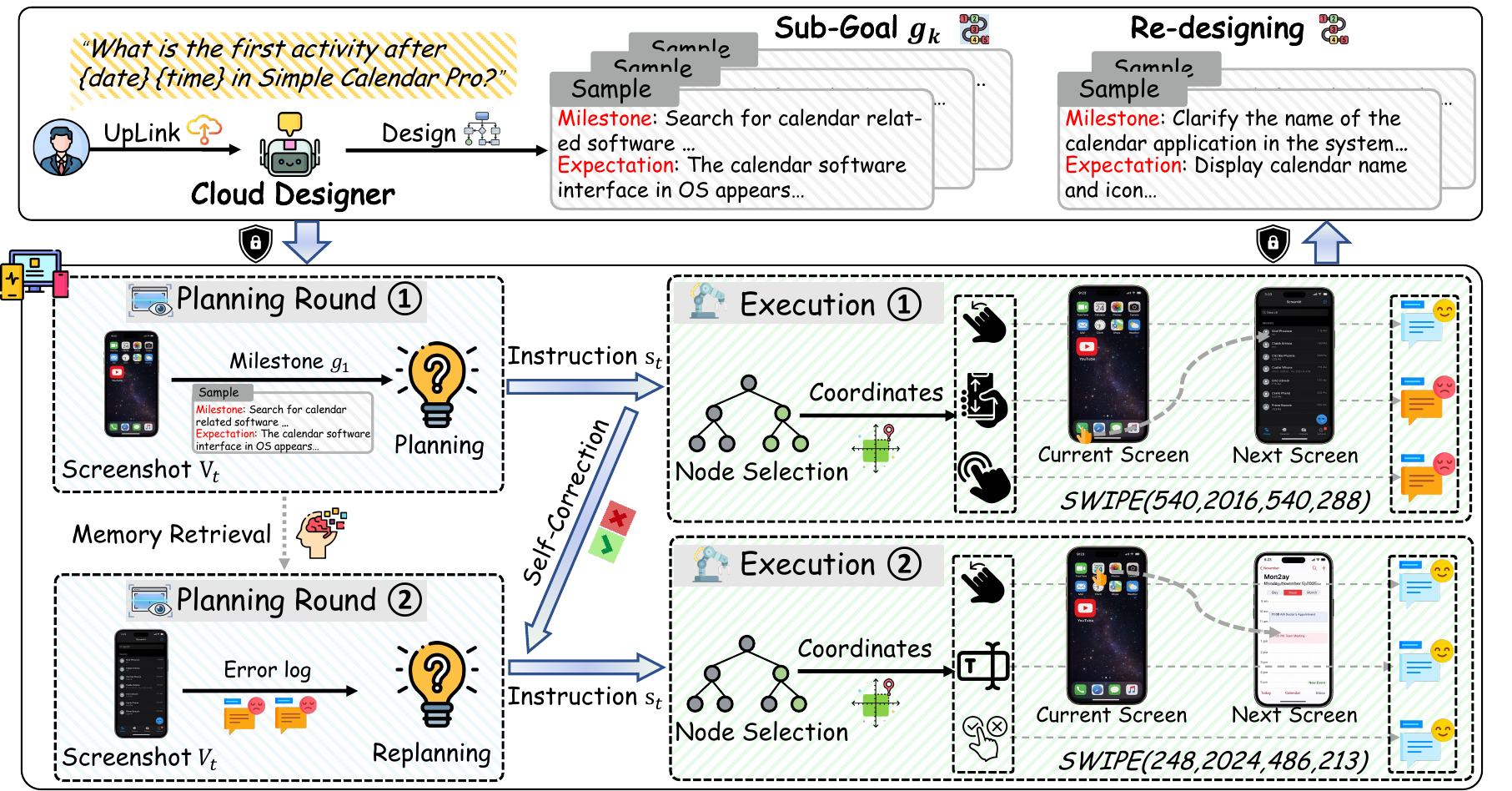}
    \vspace*{-10pt}
    \caption{\textbf{Illustration of the \textsc{AdecPilot} workflow.} The Cloud Designer orchestrates Strategic Projection, while the device-based Orchestrator and Executor collaborate to enable autonomous Self-Correction via local planning. Crucially, this closed loop is safeguarded by the HIT protocol, which enforces deterministic termination to prevent post-completion hallucinations.}
    \label{fig:workflow_illustration}
\end{figure*}

\section{Methodology}
\label{sec:method}

As shown in Fig.~\ref{fig:framework_overview}, we introduce \textsc{AdecPilot}. This framework decouples high level intent from low level grounding by assigning environment agnostic strategic design to cloud, while delegating environment specific tactical planning and execution to edge.

\subsection{Cloud-Side Strategic Design}
\label{sec:cloud_designer}

As shown in Fig.~\ref{fig:workflow_illustration}, the strategic cloud designer $\Psi_c$ functions as a high-level meta controller, operating strictly within a latent semantic space to direct the global trajectory of the task.
Unlike conventional monolithic agents that entangle strategic reasoning with heavy pixel-level processing, we implement a UI-agnostic designing mechanism driven by a text-only LLM.
Design choice is foundational to architecture: deliberately isolating designer from high dimensional raw visual stream $V_t$ and verbose view hierarchy $U_t$ compels model deriving milestones solely based upon logical reasoning and common sense knowledge regarding application workflow.

Given instruction $L_{cmd}$, failure context $H_{fail}$ (empty initially), and static functional metadata $C_m$, cloud generates $K$ coarse grained milestones directing global trajectory. Employing heuristic dispatcher, system routes instructions to task specific prompt templates via interrogative markers. Restricting $C_m$ to invariant functional schema rather than transient visual representation ensures designer remains UI Agnostic. Formal generative process:
\begin{equation}
\label{eq:cloud_projection}
G = \{ (g_k, E_k) \}_{k=1}^K \sim P_{\Psi_c}(\cdot \mid L_{cmd}, H_{fail}, C_m)
\end{equation}
Term $P_{\Psi_c}$ denotes strategic generation process executed by cloud designer $\Psi_c$. Cloud designer processes strictly abstract app metadata $C_m$ alongside user instruction $L_{cmd}$, decomposing global objective into sequence of UI agnostic milestones $G$. Initial planning enforces strictly empty history $\emptyset$. Triggering strategic redesign, cloud assimilates desensitized edge error trajectory $H_{fail}$ redefining milestones. Tuple $(g_k, E_k)$ encapsulates strategic subgoal $g_k$ and expected visual invariant $E_k$ dispatched directly to edge. By construction, cloud agent accesses zero real time rendering $V_t$, prioritizing immutable business logic over transient visual details to significantly enhance robustness across heterogeneous device form factors.


\subsection{Edge Side Collaborative Planning and Execution}
\label{sec:edge_execution}

\textbf{Insight: Reasoning Gap and Modality Mismatch.}
Design stems from dual critical observations regarding edge intelligence. First, empirical evidence reveals reasoning gap within lightweight models. Failures originate not from capacity deficit but from impulsive tendency mapping pixels directly to actions bypassing intermediate analysis. Forcing VLM to generate reasoning trace before acting significantly improves decision quality. Second, system confronts modality mismatch. While VLMs excel diagnosing dynamic visual events, models suffer high latency and low coordinate precision. Conversely, text based models operating on UI trees offer rapid structural action grounding but lack visual context handling rendering anomalies.

Resolving granularity mismatches driven by aforementioned observations, proposed bimodal collaborative architecture offloads visual processing entirely to edge. Decoupling strategic logic from implementation details minimizes latency and enhances robustness. Edge utilizes hierarchical pipeline where vision centric orchestrator conducts cognitive reasoning, guiding text centric executor through autonomous tactical planning.

\subsubsection{Orchestrator Agent: Visual Reasoning and Planning}
The execution cycle functions as a localized autonomous system driven by the Orchestrator Agent $\Phi_m$, parameterized by a quantized VLM.
Unlike conventional edge agents confined to the passive execution of atomic commands, the Orchestrator functions as a Tactical Designer.
It leverages the abstract expected state $E_k$ to perceive essential UI elements, bridging the gap between the raw screenshot $V_t$ and $E_k$.
Critically, the Orchestrator autonomously evaluates sub-task completion; if the state remains unfulfilled, it synthesizes the subsequent action based on this visual alignment analysis.
Limited edge zero-shot capabilities necessitate explicit expected states $E_k$ for effective diagnosis. To bolster robustness in non-standard UIs, we implement dynamic context injection, overriding generic priors with local logic. This process is formalized as State Alignment Optimization.
First, Orchestrator computes visual alignment score $S_t$ quantifying discrepancy between $V_t$ and $E_k$. Bypassing heuristic vector similarities, system formulates alignment as visual question answering (VQA) verification task executing natively on VLM autoregressive head. Alignment score equals conditional probability of generating affirmative indicator token given visual context and interrogative query:
\begin{equation}
\label{eq:Orchestrator_alignment}
S_t = P_{\Phi_m}(y_t = y^+ \mid V_t, \mathcal{Q}(E_k))
\end{equation}
where $y^+$ denotes affirmative vocabulary token identifying successful execution. Function $\mathcal{Q}(\cdot)$ maps abstract expected state $E_k$ into deterministic verification query. Continuous confidence measure $S_t \in [0, 1]$ dictates execution continuity. Score falling below threshold $\tau$ (empirically set to 0.85) signifies critical trajectory deviation, instantaneously triggering local tactical re-planning. The selection of $\tau$ dictates the autonomy-cost trade-off: a higher $\tau$ ensures stricter visual alignment but increases local re-planning overhead, whereas a lower $\tau$ risks grounding errors.
Instead of reporting failure to the cloud, the Orchestrator engages in planning to generate a corrective Meta-Instruction $s_t$.
This generates a local corrective trajectory without cloud intervention, achieving privacy exposure minimization and robust handling of dynamic UI elements.

\subsubsection{Executor Agent: Atomic Structural Grounding}
Upon the generation of the meta-instruction $s_t$, control is transferred to the Executor Agent $\Lambda_e$, instantiated by a lightweight text-only LLM.
Designed to alleviate the computational burden on the vision-centric Orchestrator, the Executor exploits the system's Computational Asymmetry: it operates exclusively on the textual View Hierarchy $U_t$, enabling High-Velocity Execution without pixel-level processing.
Executor treats task as dual problem integrating structural grounding and atomic actuation. Module must identify optimal DOM node $u^*$ and determine precise execution action. Replacing arbitrary heuristic matching, formulation frames grounding process as maximizing conditional semantic generation probability $P_{\Lambda_e}$. Variable $P_{\Lambda_e}$ explicitly denotes normalized sequence probability aggregating autoregressive token likelihoods generating unique identifier $ID_u$ associated with candidate node $u$ given structural view hierarchy $U_t$ and meta instruction $s_t$. Optimization objective is formalized as:
\begin{equation}
\label{eq:executor_optimization}
u^* = \operatorname*{arg\,max}_{u \in U_t^+} \left[ \log P_{\Lambda_e} \left( u \mid s_t, U_t \right) - \alpha \mathcal{R}_{struct}(u) \right]
\end{equation}
Objective balances semantic probability selecting candidate node $u$ against structural regularization term $\mathcal{R}_{struct}$ governed by scaling factor $\alpha$. Search space restricts domain to $U_t^+ = \{u \in U_t \mid v_u = 1\}$, strictly pruning non interactable elements governed by interactability indicator $v_u$ parsed natively from underlying Android structural metadata. Term $\mathcal{R}_{struct}$ mitigates visual hallucinations enforcing spatial layout constraints:
\begin{equation}
\mathcal{R}_{struct}(u) = \| \mathbf{p}_u - \mathbf{p}_{ref} \|^2
\end{equation}
Vector $\mathbf{p}_u$ denotes geometric centroid of node $u$. Spatial reference coordinate $\mathbf{p}_{ref}$ resolves location ambiguity, extracted programmatically via regex from point coordinates embedded within orchestrator textual meta instruction output. Optimization explicitly grounds abstract meta instructions into deterministic XML elements, enforcing strict structural validation and mitigating visual hallucinations~\cite{han2025swarm}.

\subsection{Hierarchical Error Recovery Mechanism}
We distinguish locally resolvable Tactical Anomalies from Strategic Failures through hierarchical control loop.

\textbf{Local Self Correction (Inner Loop).}
Addressing granularity mismatch necessitates closed feedback circuit between edge agents. Textual executor acts as rapid filter: failing to identify node $u^*$ satisfying semantic constraints triggers tactical feedback signal $\mathcal{F}_{tact}$ instead of random actuation. Visual orchestrator integrates $\mathcal{F}_{tact}$, visual context $V_t$, and expected state $E_k$ performing direct prompt conditioning. Autoregressive generation directly synthesizes revised meta instruction $s_{t+1}$ bypassing explicit probability marginalization. Paradigm proves feasibility regarding edge multi agent collaboration. Visual orchestrator and textual executor construct autonomous reasoning loop entirely on device. System resolves transient perturbations locally, strictly confining sensitive observation data to edge hardware. Design mitigates visual privacy leakage and eliminates redundant cloud token consumption.

\textbf{Strategic Redesigning (Outer Loop).}
To manage strategic deadends, system enforces tactical step budget $T_{replan}$ per sub goal. Budget serves as explicit failure exploration boundary, fundamentally decoupled from success driven HIT protocol. If orchestrator $\Phi_m$ fails to achieve expected state $E_k$ within $T_{replan}$, system synthesizes failure context $H_{fail}$. Empirical step exhaustion cases demonstrate circuit breaker operating correctly rather than algorithmic entrapment. Mechanism accommodates maximal local exploration against dynamic UI mutations, forcing edge agents to exhaust tactical possibilities before yielding. Structure completely prevents infinite cloud queries inherent to monolithic frameworks.
\begin{equation}
\label{eq:failure_context}
H_{fail} = \langle \underbrace{(g_k, E_k)}_{\text{Cloud}}, \underbrace{\{ (Q_t, a_t) \}_{t=0}^{T_{replan}-1}}_{\text{Edge}} \rangle.
\end{equation}
Variable $Q_t$ denotes textual execution trajectory of active milestone. Although $Q_t$ incurs marginal privacy exposure, textual representation significantly mitigates leakage compared to frameworks transmitting raw visual trajectories or frame summaries. Receiving $H_{fail}$, cloud designer $\Psi_c$ transitions to Diagnostician, identifying root causes and regenerating corrected trajectory $G'$. Architecture bounds error propagation, invoking expensive cloud intelligence strictly for genuine strategic failures. Maintaining optimal balance between local autonomy and global reasoning.

\definecolor{headergray}{gray}{0.92}
\definecolor{rowgray}{gray}{0.96}
\definecolor{oursblue}{HTML}{E0F7FA}
\definecolor{sectiongray}{HTML}{666666}

\begin{table*}[t]
\centering
\setlength{\aboverulesep}{0pt}
\setlength{\belowrulesep}{0pt}
\renewcommand{\arraystretch}{1.25} 

\caption{\textbf{Main Performance on AndroidWorld.} Evaluates task success rate SR, cloud token usage MT, relative cloud energy RCE. Metric RCE incorporates penalty factor $\mu=1.2$ for continuous image streaming. Symbol $^*$ indicates latest Instruct version.}
\label{tab:main_comparison}

\resizebox{\textwidth}{!}{
\begin{tabular}{llcccccc}
\toprule


\textbf{Method} & \textbf{Architecture} & \textbf{Cloud Model} & \textbf{Edge Model} & \textbf{SR}$\uparrow$ & \textbf{MT}$\downarrow$ & \textbf{RCE}$\downarrow$ & \textbf{Privacy} \\
\midrule

\multicolumn{8}{l}{\rule{0pt}{2.5ex}\textcolor{sectiongray}{\textit{\textbf{Pure Device Baselines}}}} \\

ShowUI~\cite{lin2025showui} & Single-Agent & -- & ShowUI-2B & 7.0\% & 0 & 0$\times$ & \checkmark~Safe \\

InfiGUIAgent~\cite{liu2026infiguiagent} & Single-Agent & -- & InfiGUIAgent-2B & 9.0\% & 0 & 0$\times$ & \checkmark~Safe \\

\multicolumn{8}{l}{\rule{0pt}{2.5ex}\textcolor{sectiongray}{\textit{\textbf{Pure Cloud Baselines}}}} \\

AppAgent~\cite{zhang2025appagent} & Single-Agent & GPT-4o & -- & 11.2\% & $\sim$15k & 9.0$\times$ & $\times$~High Risk \\


M3A~\cite{long2025seeing} & Multi-Agent & GPT-4o$\times$2 & -- & 28.4\% & $\sim$87k & 52.2$\times$ & $\times$~High Risk \\

\multicolumn{8}{l}{\rule{0pt}{2.5ex}\textcolor{sectiongray}{\textit{\textbf{Cloud-Device Collaborative}}}} \\

UGround~\cite{gou2025navigating} & Open-loop & GPT-4o$\times$2 & UGround-2B & 32.8\% & $\sim$45k & 27.0$\times$ & $\times$~High Risk \\


EcoAgent~\cite{yi2025ecoagent} & Closed-loop & GPT-4o & OS-Atlas-4B+Qwen2-VL-2B& 27.6\% & $\sim$3.2k & 1.9$\times$ & \textbf{!} Text Summary \\
CORE~\cite{fan2025core} & Open-loop & GPT-4o & Gemma-2-9B-IT & 26.7\% & $\sim$11.3k& 6.78$\times$ & \checkmark~\textbf{Exposure Minimization} \\
\midrule


\textbf{AdecPilot (Ours)} & \textbf{Hierarchical} & \textbf{GPT-4o} & \textbf{Qwen2.5-3B+Qwen3-VL-2B} & \underline{33.6\%}& \underline{$\sim$2k} & \underline{1.0$\times$ (Baseline)} & \checkmark~\textbf{Exposure Minimization} \\


\textbf{AdecPilot Pro (Ours)} & \textbf{Hierarchical} & \textbf{GPT-4o} & \textbf{Qwen3-4B+Qwen3-VL-2B}$^*$ & \textbf{34.5\%} & \textbf{$\sim$1.9k} & \textbf{0.95$\times$} & \checkmark~\textbf{Exposure Minimization} \\

\bottomrule
\end{tabular}
}
\end{table*}


\subsection{Adaptive Termination via Action Pruning}
\label{sec:termination}

Standard benchmarks like AndroidWorld~\cite{xu2025androidlab} impose a rigid termination tax by requiring explicit token generation. While feasible for cloud models, this protocol often induces Post-Completion Hallucination in lightweight edge models~\cite{DBLP:conf/iclr/0001S0B25}, where the agent invents destructive actions instead of stopping. To mitigate this, we propose the Hierarchical Implicit Termination (HIT) strategy, which enforces a "Fast Finish" via a multi-priority cascade restricted to the Final Milestone Phase. The execution flow is governed by three strictly ordered protocols.

Priority 1 (System Level Real Time Detection) activates entering final sub goal. System intercepts structural view hierarchy $U_t$ post atomic action capturing deterministic OS callbacks including toast notifications, triggering environment \texttt{terminate()} before model hallucinates. Priority 2 (Designer Level Logic Exhaustion) signals definitive completion triggered upon sub goal queue depletion. For question answering tasks, orchestrator asserts \texttt{ANSWER\_READY} state explicitly when visual alignment score $S_t > \tau_{qa}$ against expected text bounds, forcing immediate stop. Priority 3 (Budgetary Fallback) enforces strict global step limit $T_{max}$ resolving infinite loops. Upon triggering any priority, system wrapper executes environment \texttt{terminate()} function injecting either static success token or VLM extracted textual payload. This mechanism ensures evaluation metrics reflect agent capability rather than adherence to verbose syntax. However, HIT is calibrated for finite-horizon tasks; continuous orchestrating scenarios require adaptation to sliding-window triggers to prevent premature termination.

\section{Experiments}
\label{sec:experiments}

This section describes experimental settings in Sec.~\ref{Implementation Details}, presents quantitative comparisons between \textsc{AdecPilot} and state-of-the-art baselines across success rate and efficiency in Sec.~\ref{Analysis of Cloud Tokens and Success Rate}, analyzes privacy metrics in Sec.~\ref{Analysis of Privacy and Communication Overhead}, evaluates boundary performance in Sec.~\ref{sec:Boundary Performance and Failure Diagnostics}, and concludes with ablation studies in Sec.~\ref{Ablation Study}.

\begin{figure*}[t]
    \centering
    \includegraphics[width=\linewidth]{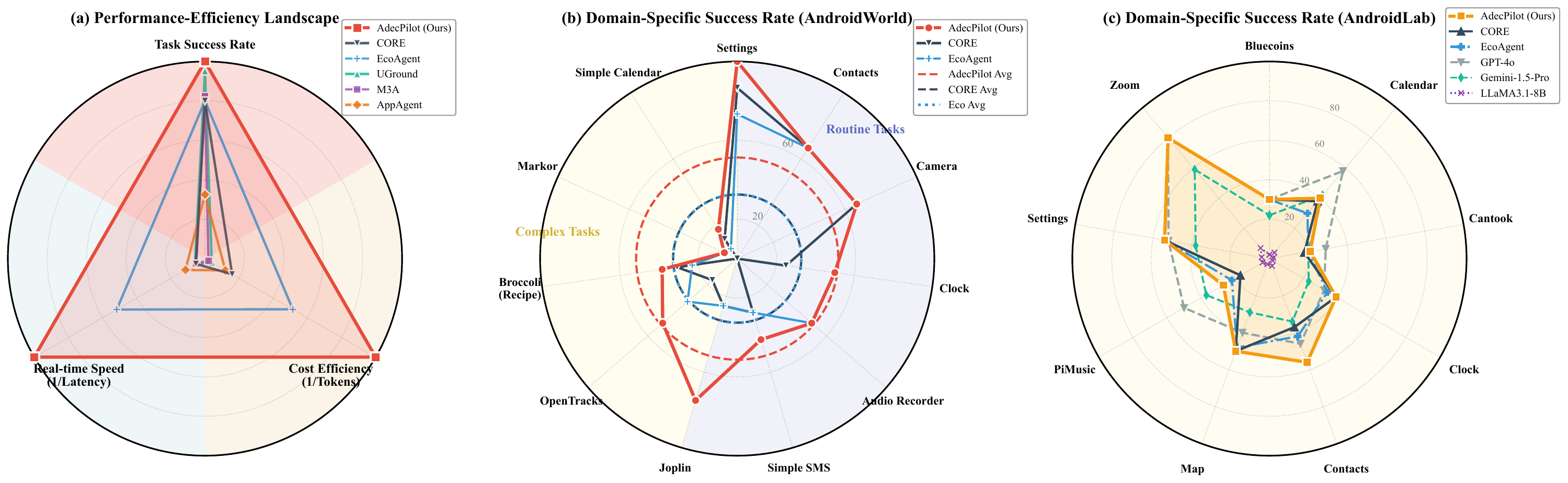}
    \vspace*{-10pt}
    \caption{\textbf{Performance analysis.} (a) Comprehensive comparison across task success rate and token cost. (b) Completion rate within AndroidWorld domain breakdown. (c) Completion rate within AndroidLab domain breakdown. Baseline selection strictly adapts to benchmark innate evaluation paradigms.}
    \label{fig:radar}
    \vspace*{-10pt}
\end{figure*}

\subsection{Implementation Details}
\label{Implementation Details}
Cloud designer $\Psi_{c}$ utilizes \textbf{GPT-4o}. Orchestrator agent $\Phi_m$ deploys quantized \textbf{Qwen3-VL-2B} model on local server equipped with NVIDIA RTX 4070 TiS GPU. Replanning limit is set to $R=1$ round. Structural regularization factor $\alpha$ within Eq.~\ref{eq:executor_optimization} is empirically set to $0.2$, balancing semantic probability against spatial constraints bypassing exhaustive ablation.

\textbf{Datasets.} Primary evaluation utilizes AndroidWorld~\cite{ICLR2025_01a83bc2} comprising 116 tasks across 20 applications. Programmatic verification ensures reproducibility over manual alternatives~\cite{wang2024mobileagentbench}. System operates on Pixel 6 emulator utilizing API 33. Accounting for environmental randomness, evaluations average across three independent runs utilizing distinct random seeds. Results exhibit minimal $\pm 0.8\%$ task success rate standard deviation, confirming objective stability and statistical significance. Although manual verification renders AndroidLab~\cite{xu2025androidlab} suboptimal for automated scaling, system adjusts execution configurations benchmarking \textsc{AdecPilot} ensuring comprehensive cross benchmark comparison.

\textbf{Metrics}. System evaluation systematically investigates four critical dimensions: efficacy, operational efficiency, latency, and privacy preservation. These are measured by following metrics.

\textit{Task Success Rate}. Metric measures efficacy, defined as percentage of successfully completed tasks relative to total test corpus.

\textit{Efficiency}. Operational efficiency is quantified via Average Cloud Calls (MC) and Cloud Token Usage (MT). To normalize resource consumption, we report Relative Cloud Energy (RCE), estimating aggregate cloud-side burden compared against baseline.

\textit{Reduction Rate}. Metric evaluates privacy mitigation measuring reduction in UI elements uploaded to cloud compared against GPT-4o baseline. Enforcing strict fairness, evaluation exclusively considers rounds where comparative methods execute identical decisions on identical UI screens~\cite{fan2025core}. Let $E_{GPT-4o}$ and $E_{ours}$ denote quantity of UI elements transmitted by GPT-4o baseline and \textsc{AdecPilot} respectively under identical conditions. Reduction rate calculation:
\begin{equation}
\label{eq:reduction_rate}
\text{RR} = \frac{E_{GPT-4o} - E_{ours}}{E_{GPT-4o}}
\end{equation}
Minimizing $E_{ours}$ inherently mitigates visual privacy leakage, objectively reducing the raw structural data exposure.

\begin{table}[h]
\centering

\caption{\textbf{AndroidWorld  Privacy Performance.} Table evaluates Success Rate(SR) and Reduction Rate(RR).}
\label{tab:privacy}

\setlength{\aboverulesep}{0pt}
\setlength{\belowrulesep}{0pt}
\renewcommand{\arraystretch}{1.2}

\begin{tabular}{@{}l c c@{}}
\toprule

\textbf{Method} & \textbf{SR} & \textbf{RR} \\
\midrule
Qwen2.5-Max (Base) & 35.3\% (41/116) & 0.0\% \\

CORE~\cite{fan2025core}(In Qwen2.5-Max) & 27.6\% (32/116) & 37.0\% \\
\midrule
\textbf{AdecPilot (In Qwen2.5-Max)} & \textbf{31.9\% (37/116)} & \textbf{75.7\%} \\
\textbf{AdecPilot Pro (In Qwen2.5-Max)} & \textbf{33.6\% (39/116)} & \textbf{79.3\%} \\
\bottomrule
\end{tabular}
\vspace{-10pt}
\end{table}

\definecolor{headergray}{gray}{0.92}
\definecolor{oursblue}{HTML}{E0F7FA}

\begin{table*}[t]
\centering
\caption{\textbf{Comprehensive Evaluation of System Overhead and Computational Efficiency.} Table compares uplink data volume, data reduction ratios, device side TFLOPs per step, and computational efficiency gains across diverse agent architectures.}
\label{tab:system_efficiency_full}

\setlength{\aboverulesep}{0pt}
\setlength{\belowrulesep}{0pt}
\renewcommand{\arraystretch}{1.3}

\begin{tabular}{@{}llcccc@{}}
\toprule
\textbf{Method} & \textbf{Edge Model Configuration} & \textbf{Uplink Data (kB)} $\downarrow$ & \textbf{Data Reduction} & \textbf{TFLOPs / Step} $\downarrow$ & \textbf{FLOPs Reduction} \\
\midrule
M3A~\cite{long2025seeing} & -- & 5831 & $1.0\times$ & 529.16 & $1.00\times$ \\
AppAgent~\cite{zhang2025appagent} & -- & 2098 & $2.8\times$ & 190.39 & $2.8\times$ \\
CORE~\cite{fan2025core} & Gemma 2 9B IT & 741 & $7.86\times$ & 138.03 & $3.8\times$ \\
EcoAgent~\cite{yi2025ecoagent} & ShowUI-2B + Qwen2-VL-2B & 120 & $48.6\times$ & 22.35 & $23.7\times$ \\
\midrule
\textbf{AdecPilot} & \textbf{Qwen2.5-3B + Qwen3-VL-2B} & \textbf{15} & \textbf{388.7$\times$} & \textbf{12.09} & \textbf{43.8$\times$} \\
\bottomrule
\end{tabular}
\end{table*}

\subsection{Analysis of Cloud Tokens and Success Rate}
\label{Analysis of Cloud Tokens and Success Rate}

Table \ref{tab:main_comparison} compares efficacy and operational cost. \textsc{AdecPilot} achieves a superior balance between execution capability and resource consumption. Notably, our method attains a success rate of 33.6\%, surpassing \textit{EcoAgent~\cite{yi2025ecoagent}} (27.6\%). Crucially, we achieve this with reduced cloud reliance. EcoAgent consumes $\sim$3.2k tokens; proposed framework requires 2k. RCE evaluation, incorporating streaming penalties, achieves 1.9$\times$ overhead reduction over strongest collaborative baseline. Against monolithic M3A at $\sim$87k tokens, RCE reduction reaches factor 52.2. Metrics validate \textit{Tactical Planning}: bimodal agents resolve granular actions without constant cloud queries. Minimal RCE 1.0$\times$ confirms offloading visual reasoning to edge, establishing standard for sustainable mobile automation.

Figure \ref{fig:radar} illustrates holistic performance efficiency trade offs and domain specific robustness. As shown in Fig. \ref{fig:radar}a, \textsc{AdecPilot} establishes optimal Pareto frontier among evaluated baselines. Monolithic cloud agents occupy low efficiency regions due to excessive token consumption. Proposed framework maximizes radar polygon area encompassing task success rate, real time speed, and cost efficiency.

\textbf{Cross Benchmark Robustness.} Fig. \ref{fig:radar}b and Fig. \ref{fig:radar}c present success rates across \textbf{AndroidWorld} categories with successful baseline cases and \textbf{AndroidLab} respectively. As illustrated visually within complex tasks region, \textsc{AdecPilot} demonstrates superior generalization in logic heavy apps including Joplin and Broccoli. Furthermore, while customized path dependent sub goals within AndroidLab interfere with autonomous edge exploration, empirical results confirm consistent superiority over monolithic and collaborative baselines under strict path constraints. As illustrated within Fig.~\ref{fig:radar}c, system achieves 56\% success rate versus CORE~\cite{fan2025core} 37\% evaluating Contacts. Unlike CORE exhibiting performance degradation in high fidelity spatial reasoning environments reaching only 17\% evaluating PiMusic, proposed bimodal architecture ensures precise Intent Grounding achieving 27\%. AndroidLab results confirm decoupling strategic intent from tactical execution effectively resolves domain specific brittle failures inherent in monolithic paradigms.

\begin{figure*}[t]
    \centering
    \includegraphics[width=\linewidth]{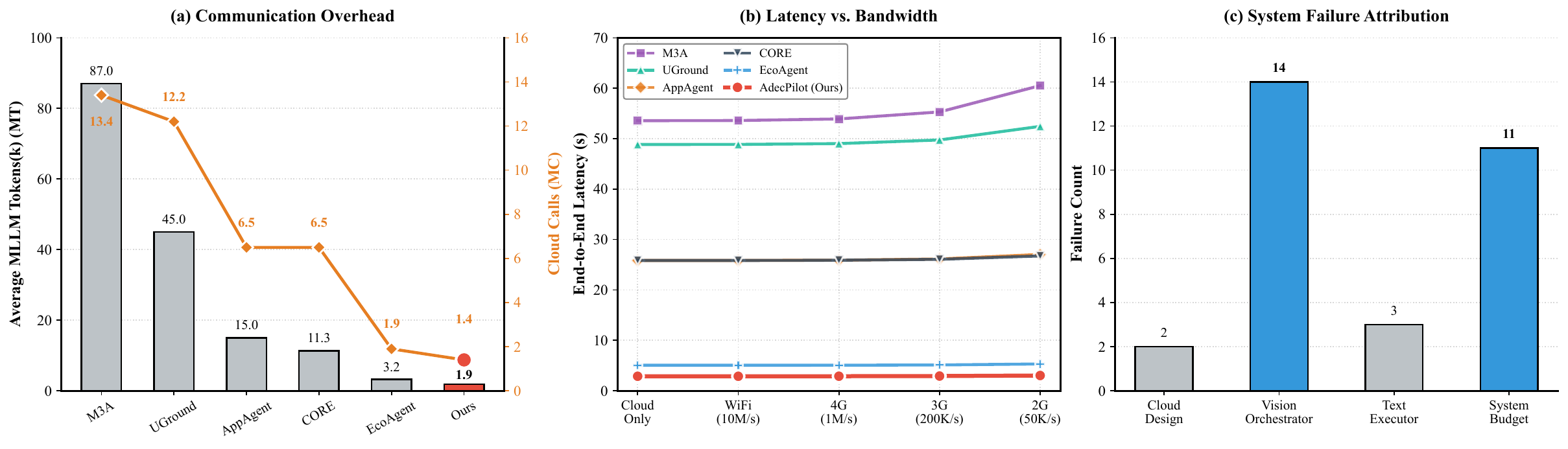}
    \vspace*{-10pt}
    \caption{\textbf{Holistic System Analysis} (a) Average cloud calls (MC) and cloud token usage (MT) comparison within AndroidWorld. (b) End to end latency comparison across weak network environments. (c) Failure case distribution analysis within AndroidWorld.}
    \label{fig:latency}
\end{figure*}

\subsection{Analysis of Privacy and Communication Overhead}
\label{Analysis of Privacy and Communication Overhead}
Table \ref{tab:privacy} reports visual privacy mitigation via Reduction Rate (RR). \textsc{AdecPilot Pro} reduces uploaded UI elements by 79.3\% over monolithic baseline; CORE achieves 37.0\%. Restricting uplink to abstract milestones confines visual data locally. Despite textual identifiers within DOM trees, decoupling mitigates image based privacy risks. \textsc{AdecPilot Pro} success rate (33.6\%) trails Qwen2.5-Max (35.3\%). Variance constitutes architectural trade off: trading 1.7\% success rate mitigates 79.3\% visual exposure while accelerating responsiveness. Strategy prioritizes operational efficiency over pure scaling~\cite{aaai/WeiD0L025}.

Table \ref{tab:system_efficiency_full} evaluates operational efficiency across uplink communication and local computation. Regarding data transmission, visual baselines including M3A~\cite{long2025seeing} incur prohibitive costs reaching 5831 kB via continuous video transmission. EcoAgent lowers overhead to 120 kB. In contrast, \textsc{AdecPilot} transmits only textual logs minimizing load to 15 kB. This represents 8.0$\times$ reduction over EcoAgent and 388.7$\times$ reduction over M3A. Quantitative evidence validates text centric collaborative approach as optimal paradigm resolving cellular transmission bottlenecks. Regarding computational efficiency, monolithic baselines including M3A demand extreme computational resources reaching 529.16 TFLOPs per step. Collaborative baseline EcoAgent executes multimodal forward passes reaching 22.35 TFLOPs per atomic action. Conversely, proposed bimodal architecture enforces computational asymmetry. Orchestrator 2B handles sparse visual alignment utilizing 1024px downsampling while Executor 3B executes precise Intent Grounding operating as text only LLM. Decoupled design reduces total per step computation to 12.09 TFLOPs. Achieving 43.8$\times$ FLOPs reduction over M3A baseline, architecture fundamentally resolves power intensive multimodal redundancy inherent within visual frameworks validating local server execution feasibility.

\textbf{Communication Efficiency.} Fig.~\ref{fig:latency}(a) illustrates average cloud calls and token consumption. \textsc{AdecPilot Pro} minimizes metrics to $1.4$ calls and $1.9$k tokens per task. Conversely, monolithic baselines exhibit severe dependence upon continuous cloud synchronization. M3A requires $13.4$ interactions consuming $87.0$k tokens, while UGround consumes $45.0$k tokens. Even optimized collaborative framework EcoAgent demands $\sim 3.2$k tokens. Superior efficiency directly stems from administrative decentralization architecture. Delegating tactical planning entirely to edge orchestrator restricts cloud interaction strictly to initial strategic decomposition and sparse failure recovery. Design eliminates redundant step validation, fundamentally resolving transmission bottlenecks.

\subsection{Boundary Performance}
\label{sec:Boundary Performance and Failure Diagnostics}

\begin{table}[t]
\centering
\caption{\textbf{Ablation Study on Architecture and Expected State.} Validates textual executor $\Lambda_e$, visual orchestrator $\Phi_m$, and explicit expected state $E_k$ within AndroidWorld dataset. Configurations utilizing Qwen2.5 3B and Qwen3 4B represent AdecPilot and AdecPilot Pro respectively.}
\label{tab:ablation_arch_state}

\setlength{\aboverulesep}{0pt}
\setlength{\belowrulesep}{0pt}
\renewcommand{\arraystretch}{1.2}

\resizebox{\linewidth}{!}{%
\begin{tabular}{@{}l c c c c@{}}
\toprule
\textbf{Method} & \textbf{SR} $\uparrow$ & \textbf{Steps} $\downarrow$ & \textbf{MT} $\downarrow$ & \textbf{Replan} $\downarrow$ \\
\midrule
w/o Executor $\Lambda_e$ & 11.2\% & 6.04 & 2492 &  0.54 \\
w/o Orchestrator $\Phi_m$ & 6.0\% & 17.94 & 6532 & 0.89 \\ 
w/o Expectation $E_k$ & 16.4\% & 13.99 & 4410 & 0.79 \\
\textbf{AdecPilot(Qwen2.5-3B)} & \textbf{33.6\%} & \textbf{9.77} & \textbf{2024} & \textbf{0.51} \\
\bottomrule
w/o Executor $\Lambda_e$ & 11.6\% & 7.02 & 2006 & 0.49 \\
w/o Orchestrator $\Phi_m$ & 7.8\% & 16.83 & 6287 & 0.85 \\
w/o Expectation $E_k$ & 19.8\% & 15.62 & 4534 & 0.83 \\
\textbf{AdecPilot Pro(Qwen3-4B)} & \textbf{34.5\%} & \textbf{10.19} & \textbf{1904} & \textbf{0.45} \\
\bottomrule
\vspace{-15pt}
\end{tabular}%
}
\end{table}

\textbf{Network Robustness.} Fig.~\ref{fig:latency}(b) depicts end to end latency across degrading bandwidth from high speed WiFi to severely constrained 2G networks. Monolithic visual streaming models including M3A and UGround suffer exponential latency surges under weak network conditions, exceeding $50$ seconds under 2G bandwidth. In stark contrast, \textsc{AdecPilot} demonstrates exceptional environmental resilience, maintaining stable $3.03$s responsiveness even under $50$K/s constraints. Architecture confines visual pixel processing and autonomous self correction loops within device hardware. Transmitting solely lightweight text based abstract milestones immunizes system against unpredictable network volatility.

\textbf{Failure Attribution.} Fig.~\ref{fig:latency}(c) objectively details system failures. Primary bottlenecks concentrate in Vision Orchestrator (14 instances) and System Budget exhaustion (11 instances). While budget exhaustion confirms designed circuit breaker prevents infinite cloud queries, non trivial proportion (11/30) indicates severe dynamic UI mutations can still entrap edge agents in excessive local replanning overhead. Orchestrator anomalies indicate edge visual verification limits remain primary capability ceiling, establishing clear target for future trajectory distillation research.

\subsection{Ablation Study}
\label{Ablation Study}
As detailed within Table~\ref{tab:ablation_arch_state}, removing textual executor $\Lambda_e$ or visual orchestrator $\Phi_m$ induces severe degradation. Omitting $\Phi_m$ collapses SR to 6.0\% while MT surges 3.2$\times$ (6532 tokens), confirming lacking visual verification triggers blind execution loops and redundant cloud redesigning. 
Expected state $E_k$ is foundational for tactical autonomy. Deleting $E_k$ forces unguided state alignment, dropping Qwen2.5 3B SR to 16.4\% elevating replan rate to 0.79, and dropping Qwen3 4B SR to 19.8\% elevating replan rate to 0.83.
Table~\ref{tab:ablation_arch_state} indicates \textsc{AdecPilot Pro} (Qwen3 4B) achieves 34.5\% SR while minimizing MT to 1904. Superior reasoning density reduces average replanning to 0.45, confirming efficient local self correction. As shown in Fig.~\ref{fig:latency}b, \textsc{AdecPilot} sustains stable 3.03s latency under 2G constraints. Unlike visual streaming baselines exceeding 50s, proposed hierarchical decoupling ensures high speed local loops remain immune to bandwidth volatility.

\textbf{Hallucination Mitigation Analysis.} Tab.~\ref{tab:ablation2} removing HIT protocol induces distinct hallucination paradigms across task categories including Question Answer and Operation. Operation tasks primarily suffer from post completion hallucination. Textual executor completes Intent Grounding but visual orchestrator lacks termination awareness. Conversely question answering tasks exhibit premature termination; agent observes target information visually but erroneously asserts completion before executing explicit textual answering. Results validate HIT protocol necessity enforcing deterministic exits and resolving category specific evaluation anomalies.

\begin{table}[t]
\centering
\caption{\textbf{Ablation Study on Hallucination Mitigation.} Total 116 AndroidWorld tasks partition into operation and question answering subsets evaluating success rate SR. Metric PCH signifies post completion hallucination rate.}
\label{tab:ablation2}
\setlength{\aboverulesep}{0pt}
\setlength{\belowrulesep}{0pt}
\renewcommand{\arraystretch}{1.2}
\begin{tabular}{@{}lcccc@{}}
\toprule
\textbf{Task type} & \textbf{Method} & \textbf{Steps} $\downarrow$ & \textbf{SR} $\uparrow$ & \textbf{PCH} $\downarrow$ \\
\midrule
\multirow{2}{*}{Operation} & w/o HIT & 13.04 & 24/91 & 11/91 \\
& \textbf{AdecPilot Pro} & 10.98 & 31/91 & 0/91 \\
\midrule
\multirow{2}{*}{Question Answer} & w/o HIT & 6.36 & 1/25 & 19/25 \\
& \textbf{AdecPilot Pro} & 7.28 & 9/25 & 3/25 \\
\bottomrule
\end{tabular}
\vspace{-10pt}
\end{table}

\section{Conclusion}
\label{sec:conclusion}

Paper introduces \textsc{AdecPilot} mobile automation collaborative framework. Administrative decentralization decouples strategic milestones from autonomous tactical planning. Architecture endows edge teams with real time visual reasoning capability resolving edge cloud cognitive lag. Bimodal design ensures robust Intent Grounding under severe network deterioration. HIT protocol eliminates hallucination via deterministic exit. Empirical results confirm framework establishes leading superiority across multiple benchmarks.

\bibliographystyle{ACM-Reference-Format}
\bibliography{Agent}

\clearpage

\appendix

\section{Hyperparameter Sensitivity Analysis}
\label{sec:appendix_hyperparameter}

Appendix investigates system robustness against critical scalar configurations. Primary focus evaluates visual alignment threshold $\tau$ governing local self correction and structural regularization factor $\alpha$ dictating spatial grounding precision.

\subsection{Visual Alignment Threshold}
Threshold $\tau$ defines autonomy boundary within visual orchestrator $\Phi_m$. Metric dictates minimum acceptable confidence score $S_t$ required asserting expected state $E_k$ completion. Table \ref{tab:hyper_tau} presents ablation across AndroidWorld dataset consistent with evaluation protocol in Section 5.1. Setting $\tau$ excessively low 0.40 permits premature milestone transition inducing cascading logic failures yielding marginal success rate 23.2\%. Conversely excessively strict configurations 0.95 trigger redundant localized replanning exhausting step budget $T_{max}$ reducing success rate 29.3\%. Empirical data validates configuration $\tau = 0.85$ establishing optimal balance maximizing success rate 34.5\% maintaining efficient average step count 10.19.

\begin{table}[h]
\centering
\caption{\textbf{Sensitivity Analysis on Visual Alignment Threshold $\tau$.} Evaluates success rate SR, average steps, and replan frequency per task. Configuration 0.85 represents baseline AdecPilot Pro.}
\label{tab:hyper_tau}
\setlength{\aboverulesep}{0pt}
\setlength{\belowrulesep}{0pt}
\renewcommand{\arraystretch}{1.25}
\resizebox{\linewidth}{!}{%
\begin{tabular}{@{}lccc@{}}
\toprule
\textbf{$\tau$} & \textbf{SR} $\uparrow$ & \textbf{Steps} $\downarrow$ & \textbf{Replan Rate} $\downarrow$ \\
\midrule
0.40 & 23.2\% & 6.88 & 0.12 \\
0.60 & 30.2\% & 7.54 & 0.23 \\
0.80 & 32.7\% & 9.35 & 0.38 \\
\textbf{0.85 (Ours)} & \textbf{34.5\%} & \textbf{10.19} & \textbf{0.45} \\
0.90 & 31.8\% & 13.45 & 0.76 \\
0.95 & 29.3\% & 16.88 & 0.88 \\
\bottomrule
\end{tabular}%
}
\end{table}

\subsection{Structural Regularization Factor}
Equation \ref{eq:executor_optimization} introduces factor $\alpha$ balancing textual semantic probability against geometric spatial constraints mitigating executor hallucination. Table \ref{tab:hyper_alpha} details spatial metric variations. Eliminating structural penalty $\alpha=0.0$ induces severe spatial hallucination rate SHR 42.1\% where executor selects semantically plausible but spatially incorrect elements. Extreme penalty $\alpha=1.0$ overrides semantic logic forcing rigid coordinate selection dropping success rate to 24.1\%. Configuration $\alpha=0.2$ optimally suppresses SHR bounding value 5.4\% sustaining maximal task completion.

\begin{table}[h]
\centering
\caption{\textbf{Sensitivity Analysis on Structural Regularization Factor $\alpha$.} Evaluates spatial hallucination rate SHR and overall success rate SR. Configuration 0.2 represents optimal baseline.}
\label{tab:hyper_alpha}
\setlength{\aboverulesep}{0pt}
\setlength{\belowrulesep}{0pt}
\renewcommand{\arraystretch}{1.25}
\begin{tabular}{@{}lcc@{}}
\toprule
\textbf{$\alpha$} & \textbf{SHR} $\downarrow$ & \textbf{SR} $\uparrow$ \\
\midrule
0.0 & 42.1\% & 22.4\% \\
0.1 & 18.5\% & 29.3\% \\
\textbf{0.2 (Ours)} & \textbf{5.4\%} & \textbf{34.5\%} \\
0.5 & 3.2\% & 30.2\% \\
1.0 & 2.1\% & 24.1\% \\
\bottomrule
\end{tabular}
\end{table}

\definecolor{headergray}{gray}{0.92}
\definecolor{rowgray}{gray}{0.96}
\definecolor{oursblue}{HTML}{E0F7FA}
\definecolor{sectiongray}{HTML}{666666}
\begin{table*}[t]
\centering
\setlength{\aboverulesep}{0pt}
\setlength{\belowrulesep}{0pt}
\renewcommand{\arraystretch}{1.25} 
\caption{\textbf{End-to-End Latency Analysis.} We report the total latency (inference + transmission) across varying network bandwidths (WiFi to 2G), where Cloud Inference time is estimated based on model modality.}
\label{tab:network_latency}
\resizebox{\textwidth}{!}{
\begin{tabular}{lcccccccc}
\toprule
\small
\textbf{Method} & 
\begin{tabular}{@{}c@{}}\textbf{MC}$\downarrow$\\[-1pt]{\scriptsize (Calls)}\end{tabular} & 
\begin{tabular}{@{}c@{}}\textbf{MT}$\downarrow$\\[-1pt]{\scriptsize (Tokens)}\end{tabular} & 
\begin{tabular}{@{}c@{}}\textbf{Cloud Latency}\\[-1pt]\end{tabular} & 
\begin{tabular}{@{}c@{}}\textbf{WiFi}\\[-1pt]{\scriptsize (10M/s)}\end{tabular} & 
\begin{tabular}{@{}c@{}}\textbf{4G}\\[-1pt]{\scriptsize (1M/s)}\end{tabular} & 
\begin{tabular}{@{}c@{}}\textbf{3G}\\[-1pt]{\scriptsize (200K/s)}\end{tabular} & 
\begin{tabular}{@{}c@{}}\textbf{2G}\\[-1pt]{\scriptsize (50K/s)}\end{tabular} & 
\begin{tabular}{@{}c@{}}\textbf{Gap(Slower)}\\[-1pt]{\scriptsize (vs. 2G)}\end{tabular} \\
\midrule

\multicolumn{9}{l}{\rule{0pt}{2.5ex}\textcolor{sectiongray}{\textit{\textbf{Pure Cloud Baselines}}}} \\
AppAgent~\cite{zhang2025appagent} & 6.46 & 15k & 25.84s & 25.85s & 25.90s & 26.14s & 27.04s & 8.9$\times$  \\
M3A~\cite{long2025seeing} & 13.39 & 87k & 53.56s & 53.60s & 53.91s & 55.30s & 60.52s & 20.0$\times$ \\
\multicolumn{9}{l}{\rule{0pt}{2.5ex}\textcolor{sectiongray}{\textit{\textbf{Cloud-Device Collaborative}}}} \\
UGround~\cite{gou2025navigating} & 12.21 & 45k & 48.84s & 48.86s & 49.02s & 49.74s & 52.44s & 17.3$\times$ \\
CORE~\cite{fan2025core} & 6.46 & 11.3k & 25.84s & 25.84s & 25.89s & 26.07s & 26.74s & 8.8$\times$  \\
EcoAgent~\cite{yi2025ecoagent} & 1.86 & 3.2K & 5.72s & 5.72s & 5.73s & 5.78s & 5.98s & 2.0$\times$ \\
\midrule
\textbf{AdecPilot Pro (Ours)} & \textbf{1.44} & \textbf{1.9k} & \textbf{2.88s} & \textbf{2.88s} & \textbf{2.89s} & \textbf{2.92s} & \textbf{3.03s} & \textbf{Baseline} \\
\bottomrule
\end{tabular}
}
\end{table*}

\subsection{End-to-End Latency and Robustness Analysis}
Table \ref{tab:network_latency} highlights a critical divergence in network resilience.
Visual baselines like \textit{M3A} are 20$\times$ slower than our approach under 2G due to image transmission, whereas \textsc{AdecPilot} maintains stable performance, rising only from 2.88s to 3.03s.
By limiting cloud interaction to abstract intent analysis which consumes 1.9k tokens against 87k, we eliminate transmission bottlenecks.
This validates that architectural decoupling renders the system immune to bandwidth volatility, ensuring responsiveness where monolithic agents fail.

\subsection{Multi User Concurrent Scalability and Power Efficiency}
\label{sec:appendix_concurrency}

Administrative decentralization fundamentally mitigates cloud computational bottlenecks during concurrent deployment. Table \ref{tab:concurrent_scalability} evaluates system scalability metrics comparing 3 concurrent mobile agents against 5 concurrent mobile agents. Monolithic baseline M3A exhibits throughput scaling increasing from 0.76 to 1.25 Queries Per Second QPS alongside Resource Consumption Energy RCE spikes reaching 258.4. Concurrent visual processing resource saturation maintains query latency high around 54.11 to 55.32 seconds. Conversely AdecPilot confines high frequency visual verification locally. Architecture delegates strictly abstract strategic planning to cloud maintaining nearly perfect linear throughput scaling from 1.44 to 2.46 QPS. System suppresses RCE below 5.0 ensuring query latency remains perfectly stable around 3.11 seconds under 5 concurrent streams.

Empirical evaluation restricts maximum concurrent instances to 5 concurrent emulated instances utilizing Pixel 6 emulator API 33 consistent with single agent evaluation protocol established within Section 5.1. Established scaling trajectories rigorously confirm edge cloud decoupling paradigm acts as prerequisite sustaining large scale mobile automation deployment. Isolated edge execution paradigm establishes optimal structural foundation deploying end cloud collaborative speculative decoding mechanisms theoretically enabling further exponential reduction regarding cloud side power consumption across massive multi user concurrent scenarios.

\begin{table*}[h]
\centering
\caption{\textbf{Cloud Infrastructure Scalability under Concurrent Loads.} Evaluates throughput Query/Sec, Relative Cloud Energy RCE, and Average Query Latency Seconds comparing 3 concurrent mobile agents versus 5 concurrent mobile agents.}
\label{tab:concurrent_scalability}
\setlength{\aboverulesep}{0pt}
\setlength{\belowrulesep}{0pt}
\renewcommand{\arraystretch}{1.25}
\resizebox{\linewidth}{!}{%
\begin{tabular}{@{}ll cc cc cc@{}}
\toprule
\multirow{2}{*}{\textbf{ }} & \multirow{2}{*}{\textbf{Method}} 
& \multicolumn{2}{c}{\textbf{Throughput(Query/Sec)} $\uparrow$} 
& \multicolumn{2}{c}{\textbf{RCE} $\downarrow$} 
& \multicolumn{2}{c}{\textbf{Query Latency(Sec)} $\downarrow$} \\
\cmidrule(lr){3-4} \cmidrule(lr){5-6} \cmidrule(lr){7-8}
& & 3 Agent & 5 Agents & 3 Agent & 5 Agents & 3 Agent & 5 Agents \\
\midrule
\multirow{3}{*}{Baselines} 
&M3A~\cite{long2025seeing} & 0.76 & 1.25  &158.1  & 258.4 & 54.11 & 55.32 \\
& UGround~\cite{gou2025navigating} & 0.72& 1.23 & 80.8 & 133.4 & 50.14 & 51.89 \\
& EcoAgent~\cite{yi2025ecoagent} &1.01& 1.62 & 5.9 & 9.1 & 5.75 & 5.96 \\
\midrule
\multirow{2}{*}{Ours} 
& \textbf{AdecPilot} & \textbf{1.44} & \textbf{2.46} & \textbf{3.1} & \textbf{4.9} & \textbf{3.06} & \textbf{3.11} \\
& \textbf{AdecPilot Pro} & \textbf{1.51} & \textbf{2.51} & \textbf{2.9} & \textbf{4.6} & \textbf{2.91} & \textbf{3.01} \\
\bottomrule
\end{tabular}%
}
\end{table*}

\section{System Prompt Formulation}
\label{sec:appendix_prompts}

Section details meta instructions governing bimodal edge cloud multi agent system. Prompts enforce strict task boundaries ensuring administrative decentralization.

\subsection{Cloud Designer Prompt}
Cloud designer objective generates UI agnostic strategic milestones avoiding low level execution coordinates. Input comprises task instruction and application metadata. Output necessitates abstract milestone formatting.

\textbf{System Formulation:} Role dictates strategic planner orchestrating mobile application workflows. Given user instruction and application metadata generate chronological milestone sequence. Strict constraint forbids coordinate generation. Milestone must encapsulate abstract functional goal. Expectation must encapsulate deterministic visual invariant indicating milestone completion. Format output utilizing tuple structure $g_k, E_k$.

\begin{promptbox}{Cloud Designer Prompt}
\textbf{role:} You are a High-Level Strategic Planner for Android automation.

\textbf{goal:} \{task\_instruction\}

\textbf{task:} Break this goal into a list of \textbf{semantic milestones}.

\textbf{for each milestone, define:}
\begin{itemize}[leftmargin=1.2em, itemsep=1pt, topsep=2pt]
    \item \texttt{instruction}: the high-level goal of this step (\textbf{what} to achieve, not \textbf{how})
    \item \texttt{expectation}: the \textbf{visual state} of the screen after this step is completed
\end{itemize}

\textbf{rules:}
\begin{itemize}[leftmargin=1.2em, itemsep=1pt, topsep=2pt]
    \item Do not specify specific UI elements such as ``click the blue button''
    \item Focus purely on the \textbf{state transition}
\end{itemize}

\textbf{good examples:}

{\ttfamily
\{"instruction": "Open the Contacts app.", "expectation": "The Contacts app main list is visible."\}
}

{\ttfamily
\{"instruction": "Fill in `Alice'.", "expectation": "The Name field shows `Alice'."\}
}
\end{promptbox}

\begin{promptbox}{Cloud Designer Prompt}

\textbf{output format:}

{\ttfamily
[

\quad \{"instruction": "...", "expectation": "..."\},

\quad ...

]
}
\end{promptbox}
\subsection{Cloud Replan Prompt}
Cloud replan objective revises the strategic milestone sequence when execution deviates from the original plan. Input comprises current task instruction, previous milestone sequence, and failed execution trajectory. Output necessitates a new milestone sequence aligned with the current environment state and remaining task objective.

\textbf{System Formulation:} Role dictates strategic replanner for mobile application workflows. Given current user instruction, previous milestones, and failure trajectory, generate a revised chronological milestone sequence that recovers from execution failure while preserving task intent. Replanning must account for completed progress, discard invalid or obsolete milestones, and infer the most plausible continuation from the current interface state. Strict constraint forbids low-level coordinates or primitive action descriptions. Each milestone must encode an abstract functional goal, and each expectation must specify a deterministic visual invariant indicating milestone completion. Format output utilizing tuple structure $g_k, E_k$.

\begin{promptbox}{Cloud Replan Prompt}
\textbf{role:} You are a High-Level Strategic Planner for Android automation.

\textbf{goal:} \{goal\}

\textbf{failed plan:}

\{old\_plan\}

\textbf{execution trace:}

\{trace\}

\textbf{task:}
\begin{itemize}[leftmargin=1.2em, itemsep=1pt, topsep=2pt]
    \item \texttt{analyze}: read the \texttt{[Orchestrator Thought]} and \texttt{[Action]} in the trace
    \item \texttt{reflect}: summarize why the previous plan failed and what should be changed
    \item \texttt{replan}: generate a new list of \textbf{high-level semantic milestones} from the \textbf{current screen}
\end{itemize}

\textbf{analysis checklist:}
\begin{itemize}[leftmargin=1.2em, itemsep=1pt, topsep=2pt]
    \item Did the agent fail to find the app?
    \item Did it get stuck on a specific screen?
    \item Did the previous instruction mislead the agent?
\end{itemize}

\textbf{recovery strategies:}
\begin{itemize}[leftmargin=1.2em, itemsep=1pt, topsep=2pt]
    \item If stuck or lost, start with \texttt{\{"instruction": "Navigate Home", "expectation": "Home screen visible"\}}
    \item If the app is not found, try a different entry point such as app drawer or app search
    \item If a step is too complex, break it into smaller milestones
\end{itemize}

\textbf{output format:}

Reflection: brief analysis of failure and recovery strategy

Plan:

{\ttfamily
[

\quad \{"instruction": "...", "expectation": "..."\},

\quad ...

]
}
\end{promptbox}

\subsection{Edge Orchestrator Prompt}
Vision orchestrator executes state alignment evaluating real time screenshot against cloud expectation $E_k$.

\textbf{System Formulation:} Role dictates visual diagnostician. Input comprises device screenshot $V_t$ and expected state description $E_k$. Evaluate visual alignment generating structured JSON payload. Formulation maps probabilistic state alignment evaluating autoregressive logits corresponding token FINISHED within status field. Detecting alignment failure generate textual meta instruction $s_t$ detailing semantic target alongside approximate spatial centroid guiding executor regularization.

\begin{promptbox}{Edge Orchestrator Prompt}
\textbf{role:} You are a strict Screen State Validator.

\textbf{current sub-goal:} ``\{sub\_goal\}''

\textbf{success criteria:} ``\{expectation\}''

\textbf{recent action history:}

\{history\}

\textbf{task:}
\begin{itemize}[leftmargin=1.2em, itemsep=1pt, topsep=2pt]
    \item \texttt{observe}: look at the screenshot and decide whether it matches the success criteria
    \item \texttt{verify history}: check whether an action was actually performed
    \item \texttt{judge}: if the screen matches expectation and history confirms action, output \texttt{FINISHED}; otherwise output \texttt{ONGOING}
\end{itemize}

\textbf{capabilities note:}
\begin{itemize}[leftmargin=1.2em, itemsep=1pt, topsep=2pt]
    \item The Executor can open app, click, type, scroll, long press, swipe, and use system keys such as back and home
\end{itemize}
\end{promptbox}

\begin{promptbox}{Edge Orchestrator Prompt}

\begin{itemize}[leftmargin=1.2em, itemsep=1pt, topsep=2pt]
    \item \textbf{Open app} is a special action: if the goal is ``Open X app'' and the app is not visible, suggest \texttt{open\_app}; do not suggest scrolling
    \item On the home screen, \texttt{scroll down} opens the app drawer, while \texttt{scroll up} opens quick settings
    \item By default, do not use \texttt{swipe} for navigation; use it only when the sub-task requires adjusting a slider
\end{itemize}

\textbf{output format:}

{\ttfamily
\{

\quad "observation": "Brief description of current screen.",

\quad "status": "FINISHED" | "ONGOING",

\quad "reasoning": "Detailed analysis of why the screen matches or fails the expectation.",

\quad "suggestion": "If ONGOING, what is the exact next move?",

\quad "spatial\_reference": "Approximate [x, y] centroid coordinate guiding textual executor."

\}

}
\end{promptbox}

\subsection{Edge Executor Prompt}
Textual executor maps orchestrator meta instruction $s_t$ against structured view hierarchy $U_t$ extracting optimal execution node.

\textbf{System Formulation:} Role dictates atomic executor. Input comprises structural view hierarchy and semantic meta instruction. Parse layout tree identifying specific interactable node $u^*$ fulfilling meta instruction intent utilizing spatial reference $p_{ref}$ enforcing structural regularization. Output requires strictly valid JSON encapsulating node ID alongside corresponding atomic action type CLICK SWIPE TYPE.

\begin{promptbox}{Edge Executor Prompt}
\textbf{role:} You are a precise UI Operator.

\textbf{current sub-goal:} ``\{sub\_goal\}''

\textbf{analysis from orchestrator agent:}

\{hint\_text\}

\textbf{spatial reference:}

\{p\_ref\}

\textbf{screen UI elements:}

\{ui\_tree\}

\textbf{task:}
\begin{itemize}[leftmargin=1.2em, itemsep=1pt, topsep=2pt]
    \item \texttt{read}: read the orchestrator analysis to understand the target element
    \item \texttt{search}: scan the screen UI element list and find the element whose text or description matches the hint
    \item \texttt{act}: select the best action using the matched element index
\end{itemize}

\textbf{available actions:}

{\ttfamily
\{"action\_type": "click", "index": <index>\}
}

Tap a UI element.

{\ttfamily
\{"action\_type": "input\_text", "text": "...", "index": <index>\}
}

Type text into a field.

{\ttfamily
\{"action\_type": "long\_press", "index": <index>\}
}

Long press an element when necessary.

{\ttfamily
\{"action\_type": "swipe", "direction": "<direction>", "index": <target\_index>\}
}

Perform a continuous physical swipe on the target element.

{\ttfamily
\{"action\_type": "open\_app", "app\_name": "..."\}
}

Directly launch an app by name.

{\ttfamily
\{"action\_type": "scroll", "direction": "<direction>"\}
}

Use \texttt{down} to scroll down or open the app drawer, and \texttt{up} to scroll up or open quick settings.

{\ttfamily
\{"action\_type": "navigate\_back"\}
}

Use the system back button.

{\ttfamily
\{"action\_type": "navigate\_home"\}
}

Go to the home screen.

\textbf{Output strictly in JSON format.}

\end{promptbox}

\section{Extended AndroidWorld Evaluation}
\label{sec:appendix_extended_results}

\begin{table*}[h]
\centering
\caption{\textbf{Granular Domain Success Rate Breakdown.} Evaluates explicit success rate across top five hardest applications within AndroidWorld.}
\label{tab:extended_domain}
\setlength{\aboverulesep}{0pt}
\setlength{\belowrulesep}{0pt}
\renewcommand{\arraystretch}{1.25}
\begin{tabular}{@{}lccc@{}}
\toprule
\textbf{Application Domain} & \textbf{AdecPilot Pro} & \textbf{CORE~\cite{fan2025core}} & \textbf{EcoAgent~\cite{yi2025ecoagent}} \\
\midrule
Calendar Tasks & \textbf{41.2\%} & 35.3\% &  29.4\% \\
Broccoli-Recipe App & \textbf{38.4\%} & 34.6\% & 30.7\% \\
Markor Editor & \textbf{28.5\%} & 25.0\% & 21.4\% \\
Simple SMS & \textbf{42.8\%} & 35.7\% & 28.6\% \\
Settings Configuration & \textbf{53.3\%} & 46.7\% & 43.3\% \\
\bottomrule
\end{tabular}
\end{table*}

Table \ref{tab:extended_domain} presents granular success rate decomposition across critical application domains within AndroidWorld extending radar chart visualization. System demonstrates profound capability processing heavily structured data apps including Simple SMS and Settings achieving success rates 42.8\% and 53.3\% respectively. Superior performance attributes directly integrating bimodal text executor retaining precise DOM parsing capabilities while visual baselines frequently misinterpret dense textual layouts.

\end{document}